\documentclass[conference]{IEEEtran}
\IEEEoverridecommandlockouts
\def\BibTeX{{\rm B\kern-.05em{\sc i\kern-.025em b}\kern-.08em
    T\kern-.1667em\lower.7ex\hbox{E}\kern-.125emX}}

\usepackage{booktabs}
\usepackage{caption}
\usepackage{tabularx}
\usepackage{multirow}

\newcolumntype{C}{>{\centering\arraybackslash}X}
\newcolumntype{L}{>{\raggedright\arraybackslash}X}

\usepackage{amsfonts}
\usepackage{bm}

\usepackage{subcaption}

\usepackage[prependcaption,textsize=scriptsize]{todonotes}
\usepackage{cite}
\usepackage{amsmath,amssymb,amsfonts}
\usepackage{algorithmic}
\usepackage{graphicx}
\usepackage{textcomp}
\usepackage{xcolor}
\usepackage{cite}
\usepackage{balance}
\usepackage{microtype}
\usepackage{url}

\usepackage[hidelinks]{hyperref}

\begin{document}

\title{Revisiting Lexicon Evaluation in \\Unsupervised Word Discovery
\thanks{Simon Malan is supported through a Google Africa PhD Fellowship as well as an NRF bursary.}}

\author{\IEEEauthorblockN{
Simon Malan,
Danel Slabbert,
Herman Kamper}
\IEEEauthorblockA{\textit{Electrical and Electronic Engineering}
\textit{Stellenbosch University},
South Africa \\
\{24227013, 24051055, kamperh\}@sun.ac.za}
}

\maketitle

\begin{abstract}
Building a lexicon from discovered word-like units is a central goal in zero-resource speech processing. 
But do our evaluations provide a trustworthy indication of lexicon quality?
A common metric, normalized edit distance, averages the phoneme edit distances between discovered units in each cluster.
We show that this metric has an inherent bias toward the quality of large clusters, inhibiting fair evaluation.
Moreover, it ignores how well true classes are distributed across clusters.
Based on established theory in clustering literature, we propose two metrics that address these shortcomings: a modified metric that weighs cluster size when assessing within-cluster consistency, and an inverse metric that assesses how true words are spread across clusters.
Through experiments on synthetic and real-world lexicons, we demonstrate that combined, these metrics are:
(1)~more closely correlated with how similar a lexicon is to the ground-truth distribution, and (2)~more robust to biases that skew lexicon evaluations.

\end{abstract}


\section{Introduction}
\label{sec:intro}

Unsupervised word discovery systems segment speech into word-like units and then cluster these units based on their similarity~\cite{park_06_wrd_disc}.
The resulting clusters form a lexicon---a vocabulary of discovered types---with each cluster ideally corresponding to a distinct word class.\footnote{Throughout, we refer to ``words'', but our work can be applied directly to evaluate systems that discover smaller units, like syllables~\cite{cho_25_sylber, baade_25_syllablelm}.}
Building high-quality lexicons is a central goal in zero-resource speech processing~\cite{ewan_20_zrc}, since it can be used for
downstream tasks such as spoken language modeling and synthesis~\cite{lakhotia_21_slm, borsos_23_synth, arora_25_slm_review}.
But determining the quality of a proposed lexicon is not trivial.

Several metrics have been proposed to evaluate general clustering problems~\cite{zhao_01_purity, rosenberg_07_vm}.
Although insightful in many applications, these metrics are not catered to lexicon evaluation.
This is because they assume that discovered units map only to a single class.
Instead, discovered units of speech can slice or span multiple ground-truth words.
The normalized edit distance (NED) metric accounts for this lexicon-specific property by transcribing units as sequences of ground-truth phonemes (obtained from forced alignments)~\cite{ludusan_14_eval}.
NED then determines the average normalized edit distance between pairs of phonemic sequences within clusters.
Other lexicon-specific evaluation metrics have also been proposed~\cite{ludusan_14_eval}, but the ease with which NED can be computed and interpreted has led to its widespread adoption~\cite{okko_15_sylseg, kamper_17_bayes_seg, kamper_17_eskmeans, okko_20_probDTW, bhati_20_se_ae, peng_22_vg_hubert, vniekerk_24_dusted, kamper_24_dpdp_hu}. However, the pairwise comparisons in NED mean that the quality of large clusters disproportionately influences the final score.
In this paper, we show that this hinders NED from fully capturing lexicon quality.

For this reason, we propose two lexicon evaluation metrics that address the cluster-size bias of NED.
Both methods ensure that each cluster's weight in the final score is proportional to its size.
The first is a weighted version of NED, while the second calculates the phonemic error rate between a cluster's modal unit and each unit in the cluster.
NED only evaluates the phonemic purity of clusters, not how well classes are grouped into distinct clusters---a property critical to lexicon quality.
To address this, we additionally propose inverse metrics, counterparts to our weighted NED and phoneme error rate metrics, that explicitly evaluate how classes are distributed over discovered clusters.
We then consider how these new forward and inverse metrics address shortcomings in lexicon evaluation, both from a theoretical and practical standpoint.

We make the following contributions.
(1) We first show that our proposed metrics meet several of the criteria proposed for clustering metrics in the machine learning literature~\cite{dom_01_clus_eval_prop, meila_03_clus_eval_prop, rosenberg_07_vm, amigo_09_clust_eval}.
Using synthetic and real-world lexicons, we then show (2) that standard NED is insufficient for lexicon evaluation, even when coupled with a metric like bitrate~\cite{ewan_20_zrc} that measures compactness; and (3) because inverse components are incorporated, our metrics more closely correlate with how similar a lexicon is to the ground truth.
We find that weighted NED and its inverse are the most comprehensive, while phonemic error~rate~and~its inverse provide faster alternatives.
Taken together, we provide metrics tailored to lexicon evaluation that are fair and easy to use.
Evaluation toolkit available at: \url{https://github.com/language-speech-learning/unit-discovery-evaluation}.

\section{Lexicons and Existing Evaluation Metrics}
\label{sec:existing}

In this section, we describe properties that well-defined metrics should fulfill, introduce common clustering evaluation metrics, and show why these are not well-suited for lexicon evaluation.
We start by defining lexicon evaluation.

An unsupervised word discovery system typically starts by dividing input speech into segments using some unsupervised method~\cite{park_08_seg_dtw, ewan_20_zrc, ankita_24_ssl_word}.
For our purposes, the exact segmentation approach is not important; we only
assume that the resulting segments, $S$, are word-like. 
(Technically they can be any length shorter than or equal to the utterance length.)
A lexicon is built by grouping these segments into $|K|$ clusters, where a cluster $k_i, i\in[1,|K|]$ contains $|k_i|\leq|S|$ discovered units.
The set of ground-truth word classes $C=\{c_1,c_2,...,c_{|C|}\}$ is defined by forced alignments, which can also be used to transcribe the discovered units; we denote this transcription as~$T$. 

\begin{figure}
    \centering
    \includegraphics[width=\linewidth]{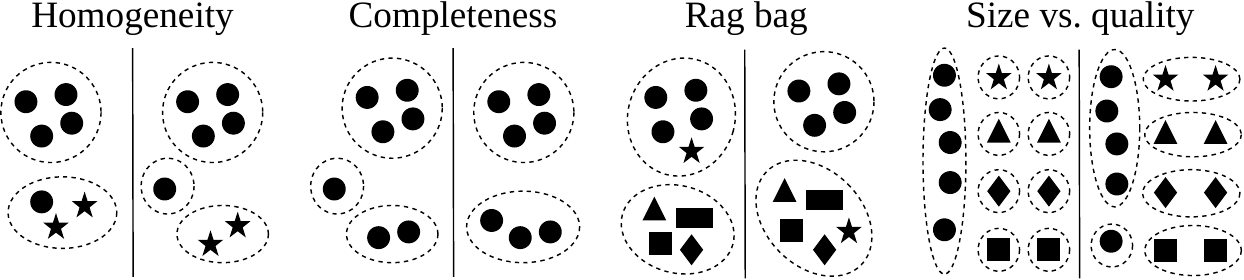}
    \caption{Four properties that a clustering evaluation metric should satisfy~\cite{amigo_09_clust_eval}. For each property, a metric should favor the lexicon on the right. Transcribed units are illustrated as filled shapes (distinct classes) and clusters as dotted regions.}
    \vspace{-2mm}
    \label{fig:properties}
\end{figure}

Many properties that evaluation metrics should possess have been proposed~\cite{dom_01_clus_eval_prop, meila_03_clus_eval_prop, rosenberg_07_vm}, but the four defined in~\cite{amigo_09_clust_eval} are the most intuitive and useful for lexicon evaluations.
As illustrated in Fig.~\ref{fig:properties}, a metric should fulfill the following properties.
\textit{Homogeneity:} clusters containing units of a single class are preferred over clusters that mix units from different classes.
\textit{Completeness:} a lexicon that splits units of the same class into as few clusters as possible is preferred.
\textit{Rag bag:} adding a unit of a new class into an inhomogeneous cluster is preferred over adding it to a perfectly homogeneous cluster.
\textit{Size vs. quality:} few clustering errors in a large class are preferred over many clustering errors in small classes.

Some properties can be exploited: assigning every unit to its own cluster (over-clustering) gives perfect homogeneity, while assigning all units to a single cluster (under-clustering) gives perfect completeness.
A metric (or combination of metrics) should penalize these extremes.
If a lexicon is perfectly homogeneous and perfectly complete, it will match the ground-truth class distribution and should be scored perfectly.

\subsection{Existing Cluster Evaluation Metrics}

Common clustering metrics typically consist of two components evaluating homogeneity and completeness, respectively.
The most basic of these is purity and inverse purity~\cite{zhao_01_purity}.
Purity counts the number of units in each cluster that match the cluster's modal class, and normalizes this count by the total number of discovered units:
\begin{equation}\label{eq:purity}
    \text{Purity}(C,K)=\frac{1}{|S|}\sum_{i=1}^{|K|}\sum_{t\in k_i}\mathbb{I}(t = c^*),
\end{equation}
with $\mathbb{I}$ the indicator for when a transcribed unit $t$ in cluster $k_i$ matches the cluster's modal class $c^*$.
Inverse purity measures completeness with a normalized count of the most common cluster assignment for each class.
Inverse purity is symmetric to~\eqref{eq:purity}, calculated by switching the $C$ and $K$ arguments. 

The v-measure~\cite{rosenberg_07_vm} is an information-theoretic metric that is the harmonic mean of two components measuring homogeneity and completeness, respectively.
Conditional entropy is used to quantify the remaining uncertainty of classes given clusters (homogeneity) and clusters given classes (completeness).
Together, the purity metrics satisfy the homogeneity and size vs.\ quality properties (but not the others), while v-measure fulfills all but the rag bag property~\cite{amigo_09_clust_eval}.

\subsection{Properties Unique to Lexicons}
\label{ssec:lex_properties}

\begin{figure}[!t]
    \centering
    \includegraphics[width=\linewidth]{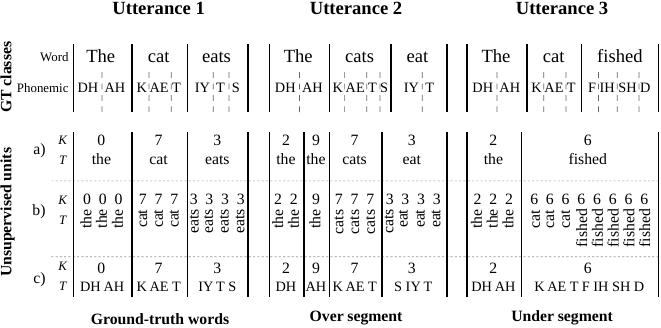}
    \caption{Three methods to transcribe discovered units: (a)~maximum-overlap, (b)~framewise, and (c)~ZeroSpeech overlap. 
    Three toy utterances are illustrated with their ground-truth (GT) classes (top) and unsupervised units (bottom).
    For each transcription method, clusters $K$ are mapped to a corresponding ground-truth class in $C$ to yield the transcription~$T$.}
    \vspace{-2mm}
    \label{fig:transcr}
\end{figure}

Although useful for general clustering evaluation, the metrics above cannot effectively be applied in lexicon evaluation due to two properties: transcriptions and unit comparisons.
To demonstrate this, we illustrate three utterances, their unit discovery segmentation, and different transcriptions $T$ in Fig.~\ref{fig:transcr}.

\textit{Transcriptions:}
Metrics like purity and v-measure assume a one-to-one mapping between units and classes.
When dealing with word discovery, these metrics typically use one of two methods to transcribe discovered units in terms of the ground truth:
Maximum-overlap (Fig.~\ref{fig:transcr}a) maps each discovered unit to its maximal overlapping class, while a framewise transcription (Fig.~\ref{fig:transcr}b) assigns a cluster-class mapping to each frame of speech.
Because the segmentation of discovered units seldom exactly match that of the ground truth, transcribing units as word classes is rarely a faithful representation of the unit's underlying acoustics.
Even when the segmentation exactly matches the ground truth (utterance~1), transcribing homophones, such as ``for'' and ``four'', results in acoustically identical units being transcribed as distinct classes.

These transcriptions have further flaws.
Maximum overlap can lead to classes being under- or over-represented: Utterance~2 in Fig.~\ref{fig:transcr}a maps both the first and second sub-word units to ``the'', while the ``cat'' class in utterance~3 is entirely skipped as the second unit maximally overlaps with ``fished''.
The framewise transcription in Fig.~\ref{fig:transcr}b addresses this issue but, in turn, is inherently weighted by the duration of each class.
Consequently, long-duration classes, like ``fished'' in Fig.~\ref{fig:transcr}, have a
disproportionate influence on the global purity score compared to shorter classes like ``the''.
Transcribing to a phonemic sequence (Fig.~\ref{fig:transcr}c) solves these issues, and is the approach followed by NED, discussed in Section~\ref{sec:ned}.

\textit{Unit Comparisons:} Transcribing units as word classes (Fig.~\ref{fig:transcr}a~and~b) reduces the evaluation logic to a binary identity check: either units match, or they do not.
In this case, similar classes such as ``cat'' and ``cats'' in Fig.~\ref{fig:transcr} are treated as completely distinct, leading to cluster~7 being evaluated as inhomogeneous.
Again, NED (below) addresses this by allowing for partial matches between phonemic sequences.

In summary, the unit-to-class mapping required for purity and v-measure are inaccurate when dealing with segmented speech units.
These metrics are therefore not appropriate for evaluating lexicons in word discovery. 
Next, we turn to existing methods that can be used for this purpose.

\section{The Normalized Edit Distance Metric}
\label{sec:ned}

In this section, we describe NED, how it fulfills the properties unique to lexicons, and explain its bias toward large clusters.
Because NED only considers homogeneity, we also discuss bitrate, which provides an indication of completeness.

NED transcribes units using what we call ZeroSpeech overlap (Fig.~\ref{fig:transcr}c)~\cite{ludusan_14_eval}.
Here, forced alignments are used to map each discovered unit to its overlapping phonemic sequence.
If a unit partially overlaps with a phoneme, the phoneme is added to the transcription if its overlap is large enough. 
This method ensures that the acoustics behind each unit, not just the word class, are fairly represented in the transcription.
For instance, in utterance~2 of Fig.~\ref{fig:transcr}c, the /S/ in ``cats'' is moved to the transcription of the last unit.
This transcription provides a realistic representation of the discovered units, corresponding to the words ``cat seat'', rather than the ground-truth ``cats eat''. 
Similarly, the ZeroSpeech transcription of utterance~3 merges the phoneme sequences of the last two ground-truth words.

To compare two phoneme sequences, NED allows partial matches by calculating the edit (Levenshtein) distance between sequences ($t,t'$), normalized by the maximum length:
\begin{equation}\label{eq:ned_clus}
    \text{NED}_i(t, t')=\frac{\text{Lev}(t,t')}{\text{max}(|t|,|t'|)}.
\end{equation} 
NED then evaluates lexicon homogeneity as the average of the normalized edit distances across all pairwise combinations of units within each cluster (lower is better):
\begin{equation}\label{eq:ned}
    \text{NED} = \frac{\sum^{|K|}_{i=1}\sum_{\{t,t'\}\subseteq k_i}\text{NED}_i(t, t')}{\sum^{|K|}_{i=1}\binom{|k_i|}{2}}, 
\end{equation}
where $\{t,t'\}$ is a unique pair of transcribed units in cluster $k_i$.

Because NED only evaluates homogeneity, bitrate~\cite{ewan_20_zrc} is typically used as a complementary metric.
The bitrate of a lexicon is the lower bound on the average number of bits required to encode the tokenized speech, expressed in bits per second (lower is better).
Therefore, bitrate does not explicitly measure completeness (it contains no information about the classes represented by the units), but rather does this implicitly by quantifying how efficiently a given lexicon allows the speech data to be stored.
We use the entropic bitrate formulation:
\begin{equation}
    \text{Bitrate}=\frac{-|S|}{\text{Dur}(\mathcal{D})}\sum_{i=1}^{|K|}\frac{|k_i|}{|S|}\text{log}_2(\frac{|k_i|}{|S|}),
\end{equation}
where Dur($\mathcal{D}$) is the duration of the data in seconds.
A tradeoff between NED and bitrate has become a standard for lexicon evaluation~\cite{malan_26_td_vs_bu}.
The degree to which NED and bitrate satisfy the clustering properties in Fig.~\ref{fig:properties} is discussed in Section~\ref{sec:theo_comp}.

\textit{Relationship to Other Lexicon Metrics:}
Other lexicon-specific metrics have also been proposed, all of which use ZeroSpeech transcriptions~\cite{ludusan_14_eval}.
Grouping precision and recall can be seen as weighted (by cluster or class size) purity and inverse purity metrics; the set of unique ZeroSpeech transcriptions ($T$ in Fig.~\ref{fig:transcr}c) are used as classes.
These can be very different from, and a much larger set than, the ground-truth set of classes (top of Fig.~\ref{fig:transcr}), meaning that grouping recall's definition of completeness does not match what is considered standard~\cite{amigo_09_clust_eval}.
Type precision and recall, respectively, measure the probabilities that the discovered units belong to the ground-truth classes and that the ground-truth classes are discovered.
However, type scores are calculated irrespective of where units are clustered, therefore none of the clustering properties in Fig.~\ref{fig:properties} are satisfied.
Furthermore, because grouping and type scores require exact matches between transcriptions, they fail to satisfy the lexicon-specific properties discussed in Section~\ref{ssec:lex_properties}.
Lastly, matching scores compare the set of discovered pairs (with substring completion) to the set of all possible ground-truth pairs.
This is comprehensive but computationally intractable for practical use in many settings because of the pairwise comparisons required.

Therefore, up to now, NED is the only theoretically viable and widely adopted metric. 
However, NED is overly biased toward the quality of large clusters, as we illustrate next.

\textit{Cluster-Size Bias:}
Because NED uses pairwise comparisons within clusters, each cluster produces $\binom{|k_i|}{2}$ normalized edit distance values within~\eqref{eq:ned}.
These pairwise comparisons provide a holistic view of a cluster's homogeneity as each unit is compared with every other unit.
NED does not aggregate edit distances per cluster; instead, it takes a global average of all pairwise edit distances.
Therefore, larger clusters have a disproportionately large effect on the overall average compared to smaller clusters.
For example, a cluster of 10 units generates 45 pairwise distances, whereas a cluster of 4 units generates only 6.
As shown in Section~\ref{sec:results}, a fully homogeneous or inhomogeneous large cluster can skew NED in either direction, resulting in a misrepresentation of the true overall homogeneity.

In summary, although NED and bitrate are often used for lexicon evaluation, NED has a cluster-size bias while bitrate does not explicitly measure completeness.

\section{Proposed Lexicon Evaluation Metrics}
\label{sec:new_metrics}

To address these shortcomings, we propose two new lexicon evaluation metrics.
Each combines a forward and inverse component that respectively evaluates homogeneity and completeness.
All our metrics use the higher-is-better convention (useful for later taking the harmonic mean).
For the sake of consistency, from this point onward we therefore use the version of NED where it is transformed into the normalized edit similarity (NES) by taking its complement: $\text{NES}=1-\text{NED}$.

\subsection{Forward Metrics}

We propose two methods to evaluate homogeneity. 
Like NES, both metrics use ZeroSpeech transcriptions (Fig.~\ref{fig:transcr}c), with comparisons based on edit distance.
However, unlike NES, each cluster's influence is directly proportional to its~size.

The first metric, weighted NES (WNES), follows the same pairwise comparison formulation as NES, but reweighs the edit distances. 
First, cluster-size bias is removed by dividing by the number of pairwise comparisons in a cluster, $\binom{|k_i|}{2}$.
Therefore, all clusters are equally weighted irrespective of their size.
But now even clusters with many units count exactly the same as clusters with a few.
What we want is a middle ground between overly-biased (NES) and equal cluster weightings.
Therefore, we reweigh the distances with the cluster size $|k_i|$.
This combined weighted distance is normalized by the total number of units (the sum of the cluster sizes):
\begin{equation}\label{eq:WNED}
    \text{WNES} = 1-\frac{\sum^{|K|}_{i=1} \frac{|k_i|}{\binom{|k_i|}{2}} \sum_{\{t,t'\}\subseteq k_i}\text{NED}_i(t, t')}{\sum^{|K|}_{i=1}|k_i|}.
\end{equation}
The result is that each unit has an equal contribution, e.g., although a cluster of size 10 will still produce 45 edit distances in~\eqref{eq:WNED}, they will be weighed as if there were only 10.

Instead of pairwise comparisons, the second forward metric, phoneme accuracy (PAcc), utilizes error rates.
Here, the~influence of each cluster's error rates naturally aligns with its size.
PAcc is calculated as the average error rate between each unit's transcription and its cluster's modal unit transcription~$t^*$: 
\begin{equation}\label{eq:PER}
    \text{PAcc}=1-\frac{1}{|S|}\sum_{i=1}^{|K|}\sum_{t\in k_i}\frac{\text{Lev}(t,t^*)}{|t^*|},
\end{equation}
where the error rate is the edit distance (Lev) normalized by the length of $t^*$.
PAcc is similar to purity in~\eqref{eq:purity}, but instead of requiring exact matches, each unit's phonemic similarity to $t^*$ is taken into account.

\subsection{Inverse Metrics}

To evaluate the completeness of a lexicon, WNES and PAcc have inverse counterparts.
Informally, these metrics determine where the ground-truth classes are clustered, and then measure how consistent the cluster assignments are within each class.

To determine where classes are clustered, we transcribe each instance of a class in terms of its cluster assignments: the sequence of clusters ($K$ in Fig.~\ref{fig:transcr}) that has a ZeroSpeech overlap with the instance.
E.g., the /DH~AH/ class in Fig.~\ref{fig:transcr} has three instances, with transcriptions $(0)$, $(2,9)$, and~$(2)$.
As in this example, the classes $C$ are defined as the set of unique phonemic realizations of words (top of Fig.~\ref{fig:transcr}).
Therefore, we consider acoustically identical words, like ``for'' and ``four'' both corresponding to /F~AO~R/, to be in the same class $c_j$.

Using these transcriptions, the inverse WNES (iWNES) metric evaluates completeness by measuring how each class is distributed over clusters.
iWNES follows the pairwise edit distance formulation of~\eqref{eq:WNED}, but now unique pairs of cluster sequences $\{y,y'\}$ belonging to class $c_j$ are compared, and weighing is according to class size:
\begin{equation}\label{eq:IWNED}
    \text{iWNES} = 1-\frac{\sum^{|C|}_{j=1} \frac{|c_j|}{\binom{|c_j|}{2}} \sum_{\{y,y'\}\subseteq c_j}\text{NED}_j(y, y')}{\sum^{|C|}_{j=1}|c_j|}.
\end{equation}

Analogously, the inverse PAcc (iPAcc) calculates the average error rate between each cluster sequence and the modal transcribed cluster sequence of its class, denoted $y^*$:
\begin{equation}\label{eq:IPER}
    \text{iPAcc}=1-\frac{1}{\sum_{j=1}^{|C|}|c_j|}\sum_{j=1}^{|C|}\sum_{y\in c_j}\frac{\text{Lev}(y,y^*)}{|y^*|}.
\end{equation}

Our inverse metrics therefore allow classes to span multiple clusters, provided the cluster sequences are consistent.

\subsection{Singleton Handling}

A lexicon-specific property that has not been discussed is singleton handling.
Singleton clusters contain only one unit, while singleton classes occur only once in the data.
We propose that correctly clustering singleton classes into singleton clusters should be credited, while clustering a non-singleton class into a singleton cluster should be penalized.
Therefore, our forward metrics assign an edit distance of 0 to singleton clusters, while our inverse metrics ignore singleton classes.
In contrast, NES ignores singleton clusters even if they contain singleton classes.

\subsection{Aggregation of Metrics}

These forward and inverse metric components can be aggregated into a single metric.
The harmonic mean of WNES and iWNES results in $F_1$-WNES.
Because PAcc and iPAcc can be negative, a harmonic mean is infeasible.
Therefore, we aggregate these metrics using their Euclidean distance from the optimal point $(1, 1)$.
We use the complement of the distance to stay consistent with our higher-is-better convention:
\begin{equation}
    d\text{-PAcc}=1-\sqrt{(1-\text{PAcc})^2+(1-\text{iPAcc})^2}.
\end{equation}

Although the separate forward and inverse metric components provide detailed insight into lexicon quality, the aggregated metrics simplify comparisons.

\section{Theoretical Comparison of Metrics}
\label{sec:theo_comp}

Before we apply our evaluation metrics to actual lexicons, we compare them to NES and bitrate on a theoretical basis.

Considering the clustering properties in Fig.~\ref{fig:properties}, both WNES and PAcc satisfy the homogeneity property.
Although NES also satisfies this property, bitrate does not---therefore the combination of these metrics contradicts one another in terms of homogeneity.
The completeness property is fulfilled by bitrate and iWNES, but not by iPAcc.\footnote{To see this, compare the RHS/LHS scores for Fig.~\ref{fig:properties} Completeness: iWNES will have scores $\frac{7}{21}$/$\frac{9}{21}$, while iPACC will have $\frac{4}{7}$/$\frac{4}{7}$.}
The rag bag property is satisfied by none of the metrics, while both the proposed inverse metrics and bitrate satisfy the size vs. quality property.

\begin{figure}
    \centering
    \includegraphics[width=0.3\linewidth]{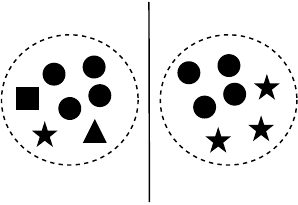}
    \caption{
    The problem of cluster matching~\cite{meila_07_matching}. A metric should evaluate the entire cluster's homogeneity, not just the number of units that match the cluster's modal class. Therefore the lexicon on the right should be preferred.}
    \vspace{-2mm}
    \label{fig:matching}
\end{figure}

Next, we consider a clustering property not present in Fig.~\ref{fig:properties}.
Although unnecessary for the comparisons thus far, the problem of matching~\cite{meila_07_matching} in Fig.~\ref{fig:matching} enables us to contrast the pairwise-comparison and error-rate methodologies of WNES and PAcc.
Specifically, the cluster-matching property applied to lexicons states that, for each cluster, non-modal units should still affect the cluster homogeneity.
Because our metrics use edit distances, the problem of matching is mitigated when units have partial matches.
However, given completely distinct units (reducing edit distances to a binary score), the error-rate formulation of PAcc only counts the number of units that match the modal unit.
The non-modal units can, in extreme cases, belong to the same or to completely distinct classes and still be scored identically.
On the other hand, the pairwise nature of NES and WNES compares all units not only to the modal unit but also to one another, providing a holistic view of cluster homogeneity.
PAcc would therefore have no preference for the two lexicons in Fig.~\ref{fig:matching}, while NES and WNES would prefer the lexicon on the right.
Class matching is the inverse of cluster matching and is analogous to the completeness property in Fig.~\ref{fig:properties}, which is fulfilled by both bitrate and iWNES but not by iPAcc.

Although WNES satisfies more properties than PAcc, the error-rate formulation of PAcc makes it much faster to compute.
Given constant edit-distance computations, the time complexity of the metrics depends on the number of comparisons made. Therefore, evaluating a cluster of size $|k_i|$ results in complexities of $\mathcal{O}(\binom{|k_i|}{2})$ for WNES and $\mathcal{O}(|k_i|)$ for PAcc.

\section{Experimental Results}
\label{sec:results}

This section examines whether the theoretical improvements of our metrics over NES and bitrate translate practically.
We consider two lexicon-evaluation experiments on LibriSpeech dev-clean~\cite{panayotov_15_librispeech} with forced alignments~\cite{mcauliffe_17_mfa} used to obtain ground-truth word classes and phonemic transcriptions.

\subsection{Real-World Lexicon Evaluation}

In this experiment, we compare evaluations using three representative real-world unsupervised word discovery systems.
Each system clusters unsupervised word-like units from~\cite{ankita_24_ssl_word} (using the hyperparameters in~\cite{malan_25_wrd_disc}) into a lexicon.
The first two systems build lexicons using the $K$-Means++~\cite{arthur_07_kmeans++} and cosine graph clustering~\cite{fortunato_10_graph_clust} methods following~\cite{slabbert_26_lex}.
To build a lexicon with less over-clustering than $K$-Means++, we use a third system, denoted $K\xrightarrow{}\text{H}$, where agglomerative hierarchical clustering~\cite{nielsen_16_hier_clust}
is used to group 1.6$(|K|)$ clusters of a trained $K$-Means++ system into $|K|$ clusters.

\begin{figure}[!b]
    \centering
    \includegraphics[width=0.9775\linewidth]{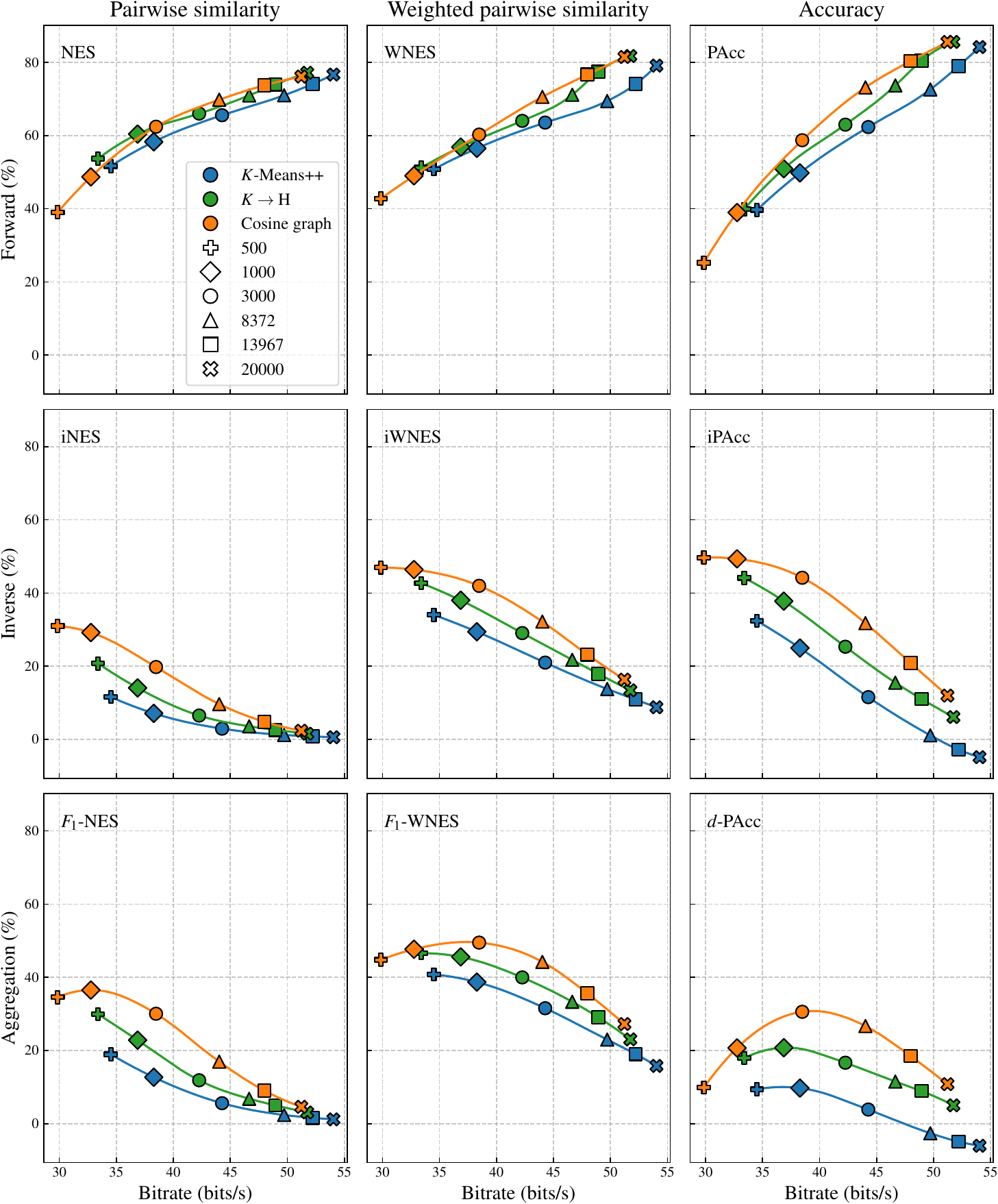}
    \caption{
    Evaluation metric scores for three unsupervised word discovery systems on LibriSpeech dev-clean.
    A range of cluster sizes $|K|$ are considered.
    Each column represents a metric family, while each row is a metric component.
    }
    \label{fig:prom-seg}
\end{figure}

In Fig.~\ref{fig:prom-seg}, each subplot uses a specific metric (top-left corner) to evaluate the three word discovery systems, each using $|K|=\{500, 1000, 3000, 8372, 13\,967, 20\,000\}$ clusters.
Here, 8372 is the number of unique phonemic realizations of words in the data, and 13\,967 is the number of clusters used by previous studies for this data~\cite{kamper_17_eskmeans}.
The values for $|K|$ lower than 8372 and higher than 13\,967 simulate under- and over-clustering, respectively.
Each row in the figure corresponds to a metric component, while each column is a family of metrics.
We plot each metric (higher is better) against bitrate (lower is better).

The top-left subplot of Fig.~\ref{fig:prom-seg} (bitrate vs. NES) is the current standard for lexicon evaluation.
Here, we choose the best lexicon based on how close it is to the optimal (top-left) point.
However, there is no obvious choice as a tradeoff between the two metrics must be made.

We can also choose a lexicon using our metrics (last two columns of Fig.~\ref{fig:prom-seg}).
The forward metrics (WNES and PAcc) show the same trends as NES (first row Fig.~\ref{fig:prom-seg}): over-clustering results in more clusters and better homogeneity.
The opposite trend holds for the inverse metrics in the last two columns of the second row: more under-clustering leads to better completeness.
Here, our inverse metrics have a strong positive correlation to bitrate, indicating that they evaluate the compactness of the lexicon while incorporating information about the underlying ground-truth classes---information unattainable by only using forward metrics.
Although a tradeoff between our forward and inverse metrics can be used to select a lexicon that best suits a specific application, we are left with a similar tradeoff as in NES vs.\ bitrate.
But maximizing an aggregated metric (last two columns, last row Fig.~\ref{fig:prom-seg}) makes this selection much easier.

\begin{figure}
    \centering
    \includegraphics[width=0.66\linewidth]{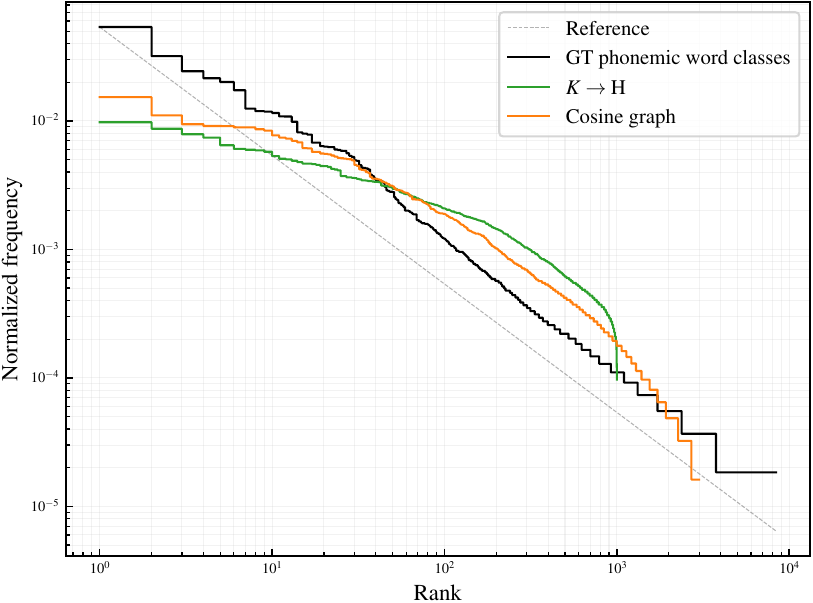}
    \caption{Normalized cluster-size distributions of the graph clustering ($|K|=3000$) and the $K$-Means++ $\xrightarrow{}$ Hierarchical ($|K|=1000$) lexicons of Fig.~\ref{fig:prom-seg}.}
    \vspace{-2mm}
    \label{fig:zipf}
\end{figure}

Based on $F_1$-WNES and $d$-PAcc, cosine graph clustering with $|K|=3000$ should be chosen as the best overall lexicon.
Looking at the standard bitrate vs.\ NES tradeoff (Fig~\ref{fig:prom-seg} top-left), this cosine graph $(|K|=3000)$ lexicon performs very similarly to the $K\xrightarrow{}\text{H}$ $(|K|=1000)$ system.
However, comparing the cluster size distribution of these systems to the ground-truth class distribution (Fig.~\ref{fig:zipf}) shows that the graph clustering system (selected as the best lexicon by our metrics) matches the ground-truth distribution much better than the $K\xrightarrow{}\text{H}$ lexicon (that could be selected using a similar NES vs.\ bitrate tradeoff).
For this reason, our metrics produce a better proxy for the similarity of a lexicon to the ground truth than NES and bitrate.
A first study to use these metrics downstream~\cite{slabbert_26_recover_zipf} show that this finding generalizes to Afrikaans and French~data.

However, why is removing the cluster-size bias required if we can add an inverse (iNES) and aggregation ($F_1$-NES) component to the NES metric family (first column of Fig.~\ref{fig:prom-seg}), leading to similar trends across lexicon-learning systems and $|K|$?
To answer this question, we construct synthetic lexicons.

\subsection{Synthetic Lexicon Evaluation}

\begin{figure}
    \centering
    \includegraphics[width=0.9775\linewidth]{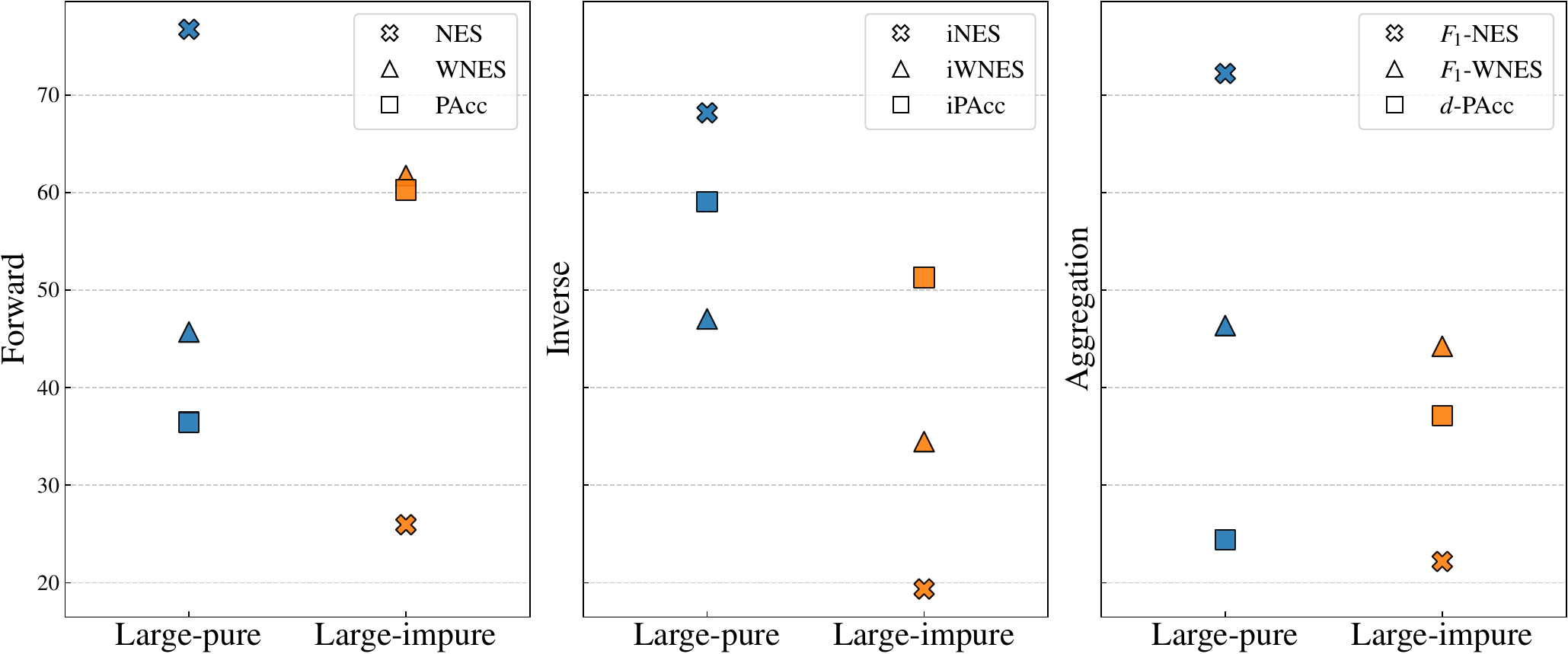}
    \caption{
    Evaluation metric scores for two synthetic lexicons built on %the 
    ground-truth classes of LibriSpeech dev-clean with $|K|\approx8372$. The large-pure lexicon has large clusters that are very pure, with smaller clusters that are less homogeneous. The large-impure lexicon is the other way around.}
    \vspace{-2mm}
    \label{fig:toy}
\end{figure}

To control the homogeneity of different clusters, we build two synthetic lexicons. 
Both cluster ground-truth word segments into $|K|\approx8372$ clusters, where the size of each cluster approximately matches that of the corresponding class, $|k_i|\approx|c_j|$.
The first lexicon, large-pure, fills each of the five largest clusters with units from distinct classes, while smaller classes are less homogeneous.
The second lexicon, large-impure, fills the five largest clusters with units from many different classes, while smaller clusters are much purer.
Although clearly different, the lexicons represent two extremes that can be seen as roughly equivalent in overall quality.
Fig.~\ref{fig:toy} shows the evaluation of both synthetic lexicons.

Comparing the results illustrates how our metrics address the cluster-size bias of NES.
Using NES, the large-pure lexicon is scored much better than the large-impure lexicon (77\% vs. 26\%), indicating that the quality of the large clusters has a disproportionate influence on NES.
The same is true for iNES, since common classes are either contained in large clusters (large-pure, 68\%) or split among smaller clusters (large-impure, 19\%).
Our forward metrics reduce the bias of NES: the large-pure system is scored much lower by WNES (46\%), while the large-impure lexicon is higher (62\%).
Although our inverse metrics show similar trends to iNES, their scores are much less exaggerated: the difference in scores between the two lexicons are 49\% for iNES and only 13\% for iWNES.

The duality between the two lexicons is balanced out in the $F_1$-WNES aggregation score.
Here, the largely equivalent (yet inverse) lexicons are evenly scored.
This result speaks to the fine-grained nature of the WNES metrics since, although PAcc and its inverse follows the same trends as WNES and its inverse, its error-rate formulation misses many non-modal comparisons. 
This does come at the cost of the WNES metrics being approximately 28 times slower to compute compared to its PAcc counterparts.
Similar to the individual forward and inverse metrics, the $F_1$-NES has a polarizing view, much preferring the large-pure~lexicon.

In summary, although the trends in Fig.~\ref{fig:prom-seg} seem similar, applying the metrics to extreme lexicons, shows that our metrics are more robust than NES (and its inverse).

\section{Summary and Conclusion}
\label{sec:conclusion}

In this paper, we revisited lexicon evaluation for unsupervised word discovery.
Although general clustering metrics provide some insight into lexicon quality, we demonstrated that tailored metrics are required for a fair assessment of a lexicon.
We showed that the most common metrics, normalized edit distance (NED) and bitrate, fail to do so: NED is overly biased toward the quality of large clusters, and bitrate does not account for information about ground-truth classes.
Our proposed metrics combine forward and inverse components that directly address 
these shortcomings.
These improvements were validated through comparisons on both real-world and synthetic lexicons.
Our metrics provided an evaluation that better captures the similarity between a lexicon and the ground truth.
We found that weighted normalized edit similarity (WNES) and its inverse are the most comprehensive metrics, while the phoneme accuracy (PAcc) and its inverse are faster to compute.
The ability of our metrics to predict downstream performance is unexplored, which will be addressed in future work.

A large body of work in zero-resource speech processing has been based on NED-based comparisons.
In light of the limitations discussed above, it is therefore difficult to assess the true extent of progress in modern unsupervised discovery systems, highlighting an important direction for future research.

\bibliographystyle{IEEEtran}
\bibliography{IEEEabrv,refs}

\end{document}